\documentclass[prd,aps,floats,twocolumn,nofootinbib,superscriptaddress,showpacs]{revtex4}
\usepackage{latexsym}
\usepackage{amssymb}
\usepackage[dvips]{graphicx}
\usepackage{amsmath}
\usepackage{amsfonts}%
\begin{document}
\title{Extended Born--Infeld Dynamics and Cosmology}

\author{M. Novello}
\affiliation{Centro Brasileiro de Pesquisas F\'\i sicas, \\Rua
Xavier Sigaud, 150, CEP 22290-180, Rio de Janeiro, RJ, Brazil}
\author{M. Makler}
\affiliation{Centro Brasileiro de Pesquisas F\'\i sicas, \\Rua
Xavier Sigaud, 150, CEP 22290-180, Rio de Janeiro, RJ, Brazil}
\affiliation{Observat\'orio Nacional - MCT, \\Rua Gal. Jos\'e
Cristino, 77, CEP 20921-400, Rio de Janeiro, RJ, Brazil}
\affiliation{Universidade Federal do Rio de Janeiro, Instituto de
F\'\i sica, \\C.P. 68528, CEP 21941-972, Rio de Janeiro, RJ, Brazil}
\author{L. S. Werneck}
\affiliation{Universidade Federal do Rio de Janeiro, Observat\'orio
do Valongo, \\Ladeira Pedro Ant\^onio, 43, CEP 20080-090, Rio de
Janeiro, RJ, Brazil}
\author{C. A. Romero}
\affiliation{Universidade Federal da Para\'\i ba, Departamento de
F\'\i sica, \\C.P. 5008, CEP  58051-970, Jo\~ao Pessoa, PB, Brazil}

\begin{abstract}
We introduce an extension of the Born--Infeld action for a scalar
field and show that it can act as unifying-dark-matter, providing
an explanation for both structure formation and the accelerated
expansion of the universe. We investigate the cosmological
dynamics of this theory in a particular case, referred as the
``Milne--Born--Infeld'' (MBI) Lagrangian. We show that this model,
whose equation of state has effectively a single free parameter,
is consistent with recent type Ia supernovae data, providing a fit
as good as for the $\Lambda$CDM model with the same number of
degrees of freedom. Furthermore, this parameter is tightly
constrained by current data, making the model easily testable with
other observables. Contrary to previous candidates for
unifying-dark-matter, the sound velocity of the MBI model is
vanishing both close to the dark matter state as well as near the
cosmological constant state. This could avoid the problems on the
matter power spectrum that were present in previous adiabatic
dark-matter/dark-energy unification models. We also present a
short discussion on the causal propagation in nonlinear scalar
field theories such as the one proposed here.

\end{abstract}

\pacs{98.80.Es, 98.80.-k, 98.80.Cq, 04.20.-q, 11.10.-z}

\maketitle

\section{Introduction}

In the current standard cosmological model, it is often assumed that
two unknown components dominate the dynamics of the universe:
dark-matter and dark-energy. Dark-matter has been for several
decades a fundamental ingredient in cosmological model building. Its
need is recognized, for instance, to explain galactic rotation
curves, virial motions and $X$-ray gas emission in galaxy clusters,
large scale flows, and gravitational lensing, all of which are
consequences of its local clustering. This clustered component
contributes to roughly a third of the critical density.

On the turn to the $21^{st}$ century, compelling evidence appeared
for a negative pressure component --- dubbed dark-energy --- capable
of producing the accelerated cosmic expansion implied by type Ia
supernovae data (SNIa, see ref. \cite{discoverySNIa}) and providing
around two thirds of the density needed to explain the almost
flatness indicated by cosmic microwave background radiation (CMBR)
observations \cite{MaxiBoom}.

Now, more evidence has been gathered to support the existence of
such a negative pressure component. For example, new SNIa
observations extending to higher redshifts \cite{Barris04,HSTSNIa}
indicate that the universe was decelerating in the past, implying
that the apparent acceleration cannot be due to simple systematic
effects on the data. Independently, constraints on the age of the
Universe \cite{age} also support the need of cosmic acceleration.
The combination of several cosmological observables, such as CMBR
anisotropies \cite{WMAP}, the large-scale structure \cite{2dFSDSS},
and $X$-ray clusters \cite{xrayclusters}, point to the same
conclusion.

These evidences were further strengthened by the reported detection
of a late Integrated Sachs-Wolfe effect through the cross
correlation of CMBR anisotropies and the galaxy distribution
\cite{GalWMAPcross}. In a nearly flat universe this effect is
significative only if some matter component with non negligible
relativistic pressure is present.

Thus, assuming the validity of Einstein's Theory of Gravity, most of
the pillars of modern cosmology seem to converge to a same
conclusion: the universe is filled with a negative pressure
component, besides dark-matter. It is nevertheless puzzling that the
dynamics of our universe be dominated by two seemingly independent
components whose nature is unknown. Although several candidates are
available for the dark-matter (such as the neutralino and the axion
\cite{DMreviews}) and for the dark-energy (such as the cosmological
constant or a scalar field \cite{DEreviews}), none has been detected
in laboratory experiments. This situation has motivated the search
for unifying-dark-matter (UDM) models, where a single component is
responsible for both structure formation and the current accelerated
expansion of the universe. 

One example of UDM is given by the Chaplygin Gas and generalizations
thereof \cite{Kamenshchik01,Billic02,bento,tese}, which were widely
studied in the literature and tested against observational data (see
for instance \cite{GCG2} and references therein). The Chaplygin gas
can be obtained from the Born--Infeld action for a scalar field and
has several physical motivations. Nevertheless, the simplest
implementation of this model is in disagreement with large-scale
structure data \cite{Sandvik}.

In this work we introduce a new model for dark-energy and
dark-matter unification based on an extension of the Born--Infeld
Lagrangian. It is shown that this matter-energy component can drive
the late time acceleration of the universe while behaving as
dark-matter in early epochs and at present in high density regions.
We focus on a subclass of such model which we refer as the
``Milne--Born--Infeld'' (MBI) Lagrangian. We study the dynamics of
the MBI model and show that a de Sitter state is an attractor for
the solutions of cosmological interest. Moreover, the ``sound
velocity'' $c_s=\delta p/\delta \rho$ is close to zero both in the
early universe and in the de Sitter epoch. This could solve the
problem of instabilities present in previous unifying-dark-matter
models \cite{Sandvik,reis04b}.

We show that cosmological observables in the MBI model, such as the
luminosity distance, depend upon a single effective free parameter
$R$ (besides the ``standard'' cosmological parameters). We use
recent data on 194 SNIa to place a strong constraint on $R$. The MBI
model provides as a good fit to the data as the $\Lambda$CDM model
with the same number of free parameters (assuming a flat universe).

The paper is organized as follows. We start by reviewing the
dynamics of purely kinetic noncanonical Lagrangians for a scalar
field and display its equivalence to a perfect fluid description
(sec. \ref{PerfectFluid}). We then discuss the effective geometry
for nonlinear scalar field theories (sec. \ref{EffGeo}) and review
the Born--Infeld model in light of this approach (sec.
\ref{BIeffGeo}). In section \ref{secEBI} we introduce an extension
of the Born--Infeld Lagrangian and discuss its role as a
unifying-dark-matter candidate. We turn to a particular case of this
Lagrangian --- the MBI model --- and derive its corresponding
equation of state in \ref{secMBI}. The sound speed and equation of
state parameter, as well as a comparison to other UDM models, are
discussed in section \ref{stability}. The dynamics of homogeneous
solutions of the MBI model is studied in section \ref{dyna}. In
section \ref{MBIaccel} it is explicitly shown that this model has a
late time accelerating phase. The Hubble parameter as a function of
the scale factor is derived in \ref{HubbleMBI} and the results are
applied to type Ia supernovae data in section \ref{SNIaMBI}. We sum
up the results and present our concluding remarks in section
\ref{conclusion}. Some speculations about the possible role of the
proposed model in the
early universe are discussed in the appendix. 

\section{Kinetic Lagrangian and the Perfect Fluid Representation\label{PerfectFluid}}

In this work we shall consider scalar field models whose Lagrangian
have a noncanonical form
\begin{equation}
L=V(\varphi)F(W)\,, \label{kLV}%
\end{equation}
where $W:=\nabla_{\mu}\varphi\,\nabla^{\mu}\varphi$.

Noncanonical (i.e., non-quadratic) kinetic terms appear rather
frequently in low-energy effective Lagrangians, generated by
radiative corrections in the underlying true theory (see e.g.
\cite{noncanonicalSUGRA}). For example, in string and supergravity
theories, nonlinear kinetic terms arise generically in the effective
action describing moduli and massless degrees of freedom
\cite{superSUSY}. Also, scalar fields are common many particle
physics theories beyond the standard model. It is therefore not
surprising that scalar fields with noncanonical kinetic terms could
play a relevant role in an effective description of nature. If we
expect the equations of motion in classical theories to be of second
order, it is natural to consider only kinetic terms which are
functions of the square of the gradient of the scalar field
\cite{Chimentok4essence}. If it is further assumed that the
Lagrangian is factorizable, such as in tachyon condensates
\cite{TachyonCond}, we are led to expression (\ref{kLV}). This type
of Lagrangian also appears as an effective theory for the low energy
expansion of fluctuations around a ghost condensate (neglecting
terms involving more than one derivative of $\varphi$, see e.g.
\cite{GhostCond}).

Lagrangians of the form (\ref{kLV}) were first introduced in the
cosmological setting in the context of inflation
\cite{kinflation,pertK}, then as a dark-energy candidate
\cite{kessence}, and finally as UDM candidates, such as in the
tachyon \cite{TachyonUDM} and ghost condensate \cite{GhostCond}
models for dark-matter and dark-energy unification.\footnote{It is
worth mentioning that other nonlinear theories, such as for spin 1
fields, may also play an important role in cosmology (see, e.g.,
ref. \cite{novello2004} and refs. therein).}

For simplicity, throughout this work we shall consider Lagrangians
with constant potentials, i.e. of the type
\begin{equation}
L=L(W)\,. \label{kL}%
\end{equation}
Such models were the first ones investigated in the inflationary
setting \cite{kinflation}. To this family belongs the Born--Infeld
action, which leads to the Chaplygin gas model. Some general
properties of this type of Lagrangian in the context of UDM are
discussed in refs. \cite{k4essence,Chimentok4essence}.

Before closing this section let us display the equivalence of
Lagrangian (\ref{kL}) to a perfect fluid description. The standard
definition of the energy-momentum tensor is given by
\begin{equation}
T_{\mu\nu}=\frac{2}{\sqrt{-g}}\,\frac{\delta(\sqrt{-g}L)}{\delta
g_{\mu\nu}}\,,
\label{PF2}%
\end{equation}
which, in the case of the scalar field with a purely kinetic Lagrangian
(\ref{kL}), leads to%
\begin{equation}
T_{\mu\nu}=-L\,g_{\mu\nu}+2L_{W}\,\nabla_{\mu}\varphi\,\nabla_{\nu}\varphi\,,
\label{PF3}%
\end{equation}
where $L_{W}:=\partial L/\partial W.$ As is well-known, any
energy-momentum tensor can be written in terms of a fluid by the
choice of a particular frame represented by an observer endowed with
a four-velocity field
$v^{\mu}$, yielding the decomposition%
\[
T_{\mu\nu}=\rho
v_{\mu}v_{\nu}-ph_{\mu\nu}+q_{(\mu}v_{\nu)}+\pi_{\mu\nu}\,,
\]
where the ten independent quantities $\rho,p,q^{\alpha}$ and
$\pi_{\alpha \beta}$ are obtained through the projections of
$T_{\mu\nu}$ onto $v^{\alpha}$ and the space orthogonal to it, and
$\pi_{\alpha\beta}$ is traceless.\footnote{Explicitly, the scalars
$\rho$ and $p$ (energy density and pressure) are defined by
$\rho:=T^{\alpha\beta}v_{\alpha}v_{\beta}$ and
$p:=-\frac{1}{3}h_{\alpha\beta}T^{\alpha\beta}$, the heat flux is
$q^{\alpha }:=h^{\alpha\beta}v^{\gamma}T_{\beta\gamma}$, and the
traceless symmetric anisotropic tension is
$\pi^{\alpha\beta}:=h^{\alpha\mu}h^{\beta\nu}T_{\mu\nu
}+ph^{\alpha\beta}$, where $h_{\mu\nu}=g_{\mu\nu}-v_{\mu}v_{\nu}$.}

In the frame commoving with the gradient of the field, defined by the
normalized vector%
\[
v_{\mu}:=\frac{\nabla_{\mu}\varphi}{\sqrt{\parallel W\parallel}}\,,
\]
the energy-momentum tensor (\ref{PF3}) is equivalent to that of a
perfect fluid, since the heat flux $q_{\alpha}$ and the anisotropic
pressure $\pi_{\mu\nu}$ vanish identically. In this case, the non
identically zero quantities are the energy density and pressure,
given by:
\begin{equation}
\rho=-L+2L_{W}W\,,\qquad p=L\,. \label{rhopL}%
\end{equation}
Moreover, since both $p$ and $\rho$ are given as a function of $W$
only, one of these relations can be inverted to provide $p=p\left(
\rho\right) $. We see therefore that there can be no bulk viscosity.
This shows that \emph{the dynamics given by any purely kinetic
Lagrangian is equivalent to that of a perfect fluid}, both in its
energy-momentum tensor and equation of state (eos).

\section{Causality and Effective Geometry\label{EffGeo}}

Motivated by the recent interest on nonlinear scalar field theories
in the cosmological context
\cite{kinflation,pertK,kessence,TachyonUDM,GhostCond,k4essence,Chimentok4essence},
we find it useful to present a brief discussion on the issue of the
causal propagation of the field.
Our purpose is to study the behavior of discontinuities of the
equations of motion around a fixed background solution. Instead of
using the traditional perturbation method (the eikonal
approximation), we shall use a more elegant formalism synthesized in
the work of Hadamard \cite{Hadamard}. In this method, the
propagation of high-energy perturbations of the field is studied by
following the evolution of the wave front, through which the field
is continuous but its first derivative is not. To be specific, let
$\sigma$ be the surface of discontinuity defined by the equation
\[
\sigma(x^{\mu})=const\,.
\]
The discontinuity of a function $J$ through the surface $\sigma$ will be
represented by $[J]_{\sigma}$, and its definition is
\[
\lbrack J]_{\sigma}:=\lim_{\delta\rightarrow0^{+}}\left(  \left.
J\right\vert _{\sigma+\delta}-\left.  J\right\vert
_{\sigma-\delta}\right)  .
\]
The discontinuities of the field and its first derivative are given by
\[
\lbrack\varphi]_{\sigma}=0\,,\qquad\lbrack\nabla_{\mu}\varphi]_{\sigma}=0\,,
\]
while for the second derivative we have
\[
\lbrack\nabla_{\mu}\,\nabla_{\nu}\,\varphi]_{\sigma}=\chi\,k_{\mu}\,k_{\nu}\,.
\]
where the vector $k_{\mu}$ is the normal to the surface of
discontinuity. Using these values in the equation of motion for the
field $\varphi$,
\begin{equation}
\nabla_{\mu}\left(  L_{W}\nabla^{\mu}\varphi\right)=0
\label{eomLw}\,,%
\end{equation}
we obtain
\begin{equation}
k_{\mu}\,k_{\nu}\,g_{eff}^{\mu\nu}=0\,, \label{C3}%
\end{equation}
where the effective metric is given by
\begin{equation}
g_{eff}^{\mu\nu}=L_{W}\,g^{\mu\nu}+2L_{WW}\,\nabla^{\mu}\varphi\,\nabla^{\nu
}\varphi\label{C4}%
\end{equation}
and $g^{\mu\nu}$ is the background metric. Only in the case of a linear theory
$L=W,$ the metric that controls the propagation of the discontinuities of the
field coincides with the background metric.

Therefore, the propagation of discontinuities of the scalar field,
which we shall refer as scalaron, follows null curves in an
effective metric that is not universal, but instead depends on the
field configuration. We should emphasize that this property is quite
generic for any nonlinear field theory (see for instance refs. \cite
{VisserEffective,NLED,EffGeoBSCG,Gibbons01} and references therein).

In terms of the background geometry we can re-write the equation
of propagation as
\begin{equation}
k_{\mu}k_{\nu}g^{\mu\nu}=-\frac{2L_{WW}}{L_{W}}\,\left(k_{\mu}\nabla^{\mu
}\varphi\right)^{2}\,. \label{kk}
\end{equation}
This means that in the background geometry the scalaron behaves as
time-like particles in cases in which $L_{W}L_{WW}<0,$ and it
behaves as tachyons in the cases in which $L_{W}L_{WW}>0$.
We shall came back to this issue in the appendix. 

\section{Exceptional Lagrangian: The Born--Infeld Case\label{BIeffGeo}}

The effective metric is an auxiliary instrument to simplify the
description of the propagation of the discontinuities of nonlinear
field theories. In nonlinear dynamics it does not correspond to the
background geometry in which the field acts. From this property one
could conclude that such geometry should not appear in any aspect of
the dynamics of the field. This is indeed the case in the most
common theories, however, in certain nonlinear theories the dynamics
is described in terms of quantities associated to the effective
metric. For example, in Electrodynamics there is at least one
exceptional case which is worth to point out: the nonlinear field
theory proposed by Born and Infeld \cite{BI}. In that theory, the
Lagrangian is written in terms of the determinant of the effective
metric $L=(-\det\,g_{eff})^{1/4}$ \cite{NovelloEffective}. We shall
show below that a similar situation occurs for the nonlinear scalar
field theory. In order to describe such special dynamics we must
face the question: what is the property of the Lagrangian $L(W)$
such that its corresponding dynamics is given only in terms of the
effective metric? We shall follow the same lines as for the case of
Electrodynamics \cite{NLED}.

We note that the (inverse) covariant metric tensor of the effective metric
(\ref{C4}), defined by the relation $g_{\mu\nu}^{eff}g_{eff}^{\nu\lambda
}=\delta_{\mu}^{\lambda}$, is given by
\begin{equation}
g_{\mu\nu}^{eff}=\frac{1}{L_{W}}\,g_{\mu\nu}-\frac{2L_{WW}}{L_{W}%
\,(L_{W}+2WL_{WW})}\,\nabla_{\mu}\varphi\,\nabla_{\nu}\varphi\,. \label{EL1}%
\end{equation}
A straightforward calculation shows that the evaluation of the determinant of
the effective metric yields the result:
\begin{equation}
\det\,g_{eff}^{\mu\nu}=L_{W}^{3}\,(L_{W}+2W\,L_{WW})\,. \label{detg}%
\end{equation}
It then follows that the unique theory that can be written in terms
of its associated effective metric is the one provided by the
Born--Infeld like form:
\begin{equation}
L_{BI}=-\sqrt{bW+e}\,, \label{BILag}%
\end{equation}
where $b$ and $e$ are constants. Indeed, using (\ref{BILag}) into
the expression of the determinant (\ref{detg}) we obtain
\begin{equation}
L_{BI}=const.\,\left(\det\,g_{\mu\nu}^{eff}\right)^{1/6}. \label{C7}%
\end{equation}

The energy density and pressure of the effective fluid description
for the
Born--Infeld dynamics (from eqs. \ref{BILag} and \ref{rhopL})%
\begin{equation}
\rho=e/\sqrt{bW+e}\,,\qquad p=-\sqrt{bW+e}\,, \label{rhopChap}%
\end{equation}
are such that the equation of state takes a very simple
form\footnote{It is worth noting that, although Lagrangian
(\ref{BILag}) has two free parameters, only one
appears in the eos.}%
\begin{equation}
p=-\frac{e}{\rho}\,. \label{ChapEOS}%
\end{equation}
A fluid with the above eos is known as Chaplygin gas\footnote{This
name, which became popular in the $d$-brane setting, is off course
\emph{inspired} on the eos proposed by Chaplygin in 1904. In the
original expression, $\rho$ is only the internal energy, not the
total relativistic one, and the weak energy condition is not
violated, as opposed to what is needed in the cosmological case.}
\cite{Chap}.

The Born--Infeld action for a scalar field was first introduced by
Heisenberg in 1939, in the context of meson theory \cite{Heis}. This
model reappeared in the string theory setting, where the
Born--Infeld Lagrangian is associated to the parametrization
invariant Nambu--Goto $d$-brane action in a ($d+1,1$) space-time
\cite{Jackiw99}. The Chaplygin gas has recently attracted much
interest in the cosmological framework, both as a candidate for
dark-energy \cite{Kamenshchik01,BeanDore} as well as for
unifying-dark-matter \cite{Billic02,GCG4ess}. In fact, as can be
seen from (\ref{ChapEOS}) it has a negative pressure that could
power the accelerated expansion of the universe. Also, the pressure
is negligible for high density regions and ``mechanical''
perturbations are stable ($dp/d\rho=c_{s}^{2}>0$). Therefore this
fluid could also be responsible for structure formation. Because of
these properties and its simplicity, equation (\ref{ChapEOS}) arises
naturally as a toy model allowing dark-energy and dark-matter
unification \cite{tese}. The behavior of the Chaplygin gas, and
simple generalizations thereof, as unifying-dark-matter (or simply
quartessence\footnote{As in this scenario there is a single extra
component besides ``baryons'', neutrinos, and radiation (as opposed
to having both dark-matter and dark-energy), the
unifying-dark-matter is dubbed as quartessence \cite{GCG1}.}) was
extensively discussed in the literature (see, for instance, refs.
\cite{tese,Billic02,bento,GCG1,Sandvik,CMBGCG,GCG2,GCG4ess,skewness,reis03a,ZhuGCG}
and references therein).

\section{Extended Born--Infeld Lagrangian\label{secEBI}}

As pointed out in the previous section, the Born--Infeld Lagrangian
arises rather naturally in the context of a geometric description
for nonlinear scalar fields. Moreover, as mentioned above, besides
its formal appeal, this theory has an interesting behavior in the
cosmological context. In fact, eos (\ref{ChapEOS}) has provided the
first model of dark-matter and dark-energy unification
\cite{tese,Billic02}. Nevertheless, this model is ruled out by
observations of weak gravitational lensing combined with the
large-scale galaxy power spectrum  \cite{Sandvik} and by the
anisotropies on the CMBR \cite{CMBGCG}. Besides, it is also
disfavored by a combination of observational data on the background
evolution of the universe \cite{GCG2,ZhuGCG}.

A generalization of the Born--Infeld Lagrangian was proposed in
\cite{bento} and leads to the so-called \emph{Generalized Chaplygin
gas} \cite{Billic02}. However, this model suffers from the same
problems as the Chaplygin gas itself \cite{Sandvik,CMBGCG}.

Nevertheless, the successful aspects of the Born--Infeld model as
a candidate for dark-energy and dark-matter unification motivate
the search for other models that still retain some of its
simplicity at the same time avoiding its shortcomings. As shall be
discussed in details in the forthcoming sections, such a model
could be given by the Extended Born--Infeld (hereafter EBI)
Lagrangian represented by
\begin{equation}
L=-\sqrt{aW^{2}+bW+e}\,. \label{EBILag}%
\end{equation}

The original idea of Born and Infeld \cite{BI} was to deal with a
theory that describes the evolution of electromagnetic fields that
are bounded from above. Here we apply such limitation idea to the
case of a scalar field, as in Lagrangian (\ref{BILag}). However,
instead of having only a maximum value we also require the existence
of a minimum (such that $a<0$ and $b^{2}>-4ae$). Therefore, $W$ is
defined between the two roots of the polynomial in (\ref{EBILag}).
From Lagrangian (\ref{EBILag}), the energy density and pressure of
the effective fluid are given by
\begin{align}
\rho &  =-L+\frac{W\,(b+2aW)}{L}\,,\label{rhoEBI}\\
p  &  =L\,. \label{pEBI}%
\end{align}

The equation of motion (\ref{eomLw}) derived from Lagrangian
(\ref{EBILag}) is
\begin{equation}
\nabla_{\mu}\left(\frac{(b+2aW)}{L}\,\nabla^{\mu}\varphi\right)=0\,.
\label{PF6}%
\end{equation}
Therefore, a particular solution for the dynamics of the scalar field is given
by a constant value of $W$, namely
\begin{equation}
W_{0}=-\frac{b}{2a}\,. \label{FS1}%
\end{equation}
In this case, the functional derivative of the Lagrangian $L_{W}$
vanishes and consequently the tensor $T_{\mu\nu}$ becomes identical
to the energy distribution of a cosmological constant.
We shall call such a configuration the \emph{fundamental state}. As
we shall see later on, in the cosmological case there exists two de
Sitter states associated with such constant $W_{0}$, one which is
stable and the other unstable. Equation (\ref{PF6}) allows also the
trivial solution $\varphi=const$. which corresponds to another
``vacuum'' state. We shall see latter that, at least in the
cosmological case, the physically relevant vacuum is given by the
fundamental state (\ref{FS1}).

Notice that, near the roots of the quadratic function, $L\simeq0$
and the scalar field acts as pressureless matter.
Besides, close to at least one of these roots the density is
positive. Thus we have the possibility of a matter-energy component
that interpolates between dust (close to a root of \ref{EBILag}) and
a cosmological constant like configuration (at the fundamental
state). Moreover, the presence of $L$ in the denominator in
expression (\ref{rhoEBI}) shows that the scalar field behaves as
dust in high density regions. This situation is typical from
unifying-dark-matter models, where this component acts as
dark-matter in clustered regions and as dark-energy in low density
regions, providing and explanation for the accelerated expansion of
the universe \cite{tese}. We shall investigate this model in the
cosmological setting in the forthcoming sections. Our analysis will
be focused on a special case of Lagrangian (\ref{EBILag}) as
described below.

\section{The Milne--Born--Infeld Lagrangian\label{secMBI}}

In the reminder of this article we shall restrict ourselves to a
subclass of Lagrangian (\ref{EBILag}), namely the case in which
$e=0$. In this situation one root corresponds to a $p=\rho=0$ state
and there is a positive root (we choose $b>0$) corresponding to a
dust limit. As we shall see in section \ref{dyna} this model allows
a solution representing an expanding empty ($\rho=0$) universe,
which is known as the Milne solution. For this reason
we refer to this model as the ``Milne--Born--Infeld'' (hereafter
MBI) Lagrangian. In this case it is convenient to express
(\ref{EBILag}) as
\begin{equation}
L=-A\sqrt{W\left(\Sigma^{2}-W\right)}\,, \label{MBILag}%
\end{equation}
where $A:=\sqrt{\left\vert a\right\vert}$ and $\Sigma^{2}:=bA^{-2}$.
For this model, the whole admissible spectra of values for $W$ is
contained in the domain $(0,\Sigma^{2})$. The fundamental state is
located at $\Sigma^{2}/2$.

In the MBI model, the energy density and pressure (eqs. \ref{rhoEBI} and
\ref{pEBI}) are given by
\begin{align}
\rho &  =\frac{AW^{2}}{\sqrt{W\left(\Sigma^{2}-W\right) }}\,,\label{rhoMBI}\\
p  &  =-A\sqrt{W\left(\Sigma^{2}-W\right)}\,. \label{pMBI}%
\end{align}
Here the expression of the eos $p\left(\rho\right)$ is rather
complicated, as opposed to the Born--Infeld case (eq.
\ref{ChapEOS}). It is more convenient to write $\rho$ as a function
of $p$:
\[
\rho=\left(A\Sigma^{2}+\sqrt{A^{2}\Sigma^{4}-4p^{2}}\right)^{2}\!/\,4p\,.
\]
It is easy to see that, for high densities ($\rho\gg A\Sigma^{2}$),
the equation of state goes asymptotically to the Chaplygin one
\[
p\simeq-\frac{A^{2}\Sigma^{4}}{\rho}\,.
\]
The density and pressure of the fundamental state are given by
\begin{equation}
\rho_{\mathrm{v}}:=\frac{A\Sigma^{2}}{2}=-p_{\mathrm{v}}\,. \label{rhov}%
\end{equation}

\begin{figure}[ptb]
\centering \hspace*{0.in}
\includegraphics[height= 8.0 cm,width=8.5cm]{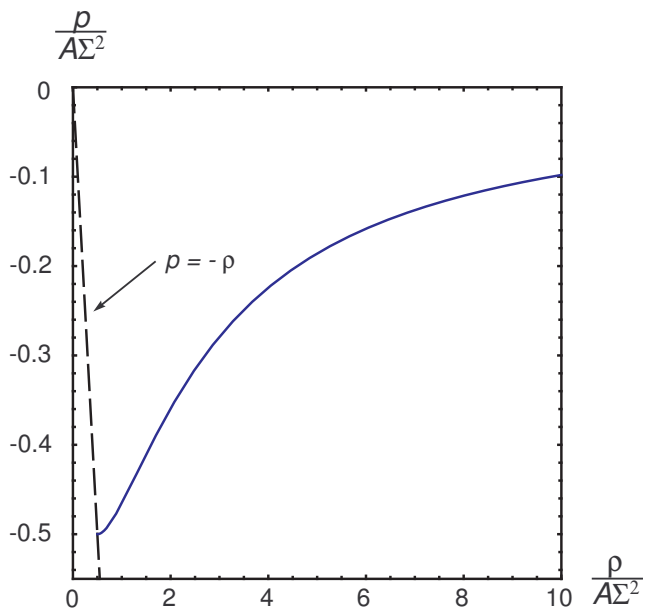}
\caption{Equation of state of the MBI model. The dimensionless
pressure $p/\!\left(A\Sigma^{2}\right)$, is plotted as a function of
$\rho/\!\left(A\Sigma^{2}\right)$ starting from the fundamental
state $\rho_{\mathrm{v}}= A\Sigma^{2}/2$. Also displayed is the
``vacuum'' eos (dashed curve).}
\label{MBIeos}%
\end{figure}

In figure \ref{MBIeos} we show $p$ as a function of $\rho$ in the
interval $\Sigma^{2}>W>\Sigma^{2}/2$, from the dust to the
fundamental state. As shall be shown in section \ref{dyna}, this is
the relevant interval for cosmology.

Before exploring the dynamics given by the MBI Lagrangian
(\ref{MBILag}), let us point out a fundamental difference with
respect to the Born--Infeld case and the Extended--Born--Infeld with
$e\neq0$. In these two models, the Lagrangian has a canonical form
for small values of $W$ ($W\ll e/b$ and $W\ll b/a$). However, the
MBI model does not have such a canonical limit. Nevertheless, we
shall see that the MBI Lagrangian provides an interesting model for
the dark sector of the universe.

\section{Sound Speed and Equation of State Parameter\label{stability}}

In the fluid interpretation, the important quantity that controls
the level of stability of perturbations is the ``sound velocity''
squared $c_{s}^{2}=\left(\partial p/\partial W\right)
/\left(\partial\rho/\partial W\right)$ \cite{pertK}, which in the
MBI model is given by
\begin{equation}
c_{s}^{2}=\frac{2W^{2}-3\Sigma^{2}W+\Sigma^{4}}{2W^{2}-3\Sigma^{2}W}\,.
\label{TS1}%
\end{equation}
In the interval $\Sigma^{2}/2<W<\Sigma^{2}$ (from the dust to the de
Sitter state) it is immediate to see that the quantity $c_{s}^{2}$
is always positive. Its maximum is given by $c_{s\max}^{2}=1/9$ and
occurs at $W_{a}=3\,\Sigma^{2}/4$, the value at which the scalar
field's self gravity starts being repulsive (see section
\ref{MBIaccel}). For $W<\Sigma^{2}/2$, $c_{s}^{2}<0$ and the fluid
is mechanically unstable. However, as we shall see in section
\ref{dyna}, this is outside the physical domain and such values are
never attained in the cosmological framework.

\begin{figure}[ptb]
\centering \hspace*{0.in}
\includegraphics[height= 8.0 cm,width=8.5cm]{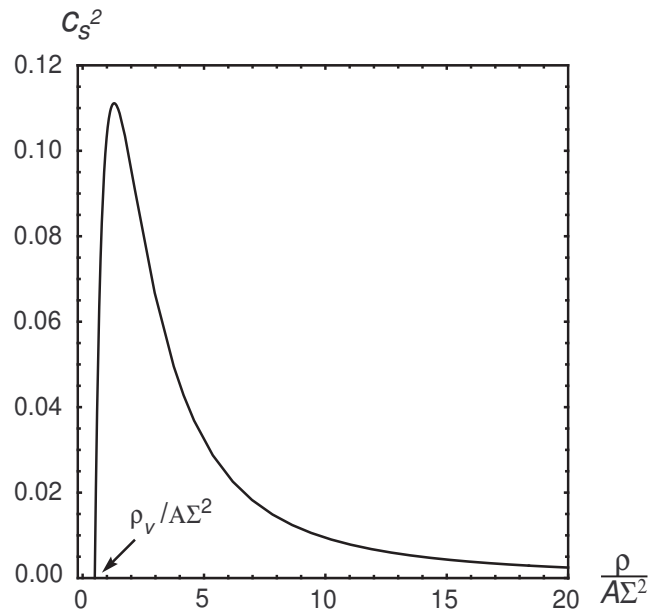}
\caption{The sound velocity squared as a function of the
dimensionless density $\rho/\left(A\Sigma^{2}\right)$ in the MBI
model.} \label{MBIcs}
\end{figure}

In figure \ref{MBIcs} we plot the sound velocity as a function of
the density starting at the fundamental state (\ref{rhov}). It is
interesting to point out some features of the MBI model regarding
the sound velocity. First, its maximum value is a third of the
velocity of light, whereas in the Born--Infeld model it is equal to
$c.$ Most importantly, near the de Sitter state the sound velocity
is zero. As shall be discussed later, our universe is currently
approaching this de Sitter state, thus one might expect the present
scalaron sound velocity to be small. This is in contrast to most
quartessence models studied so far in the literature, where the
maximum sound velocity is attained precisely at the de Sitter state
\cite{reis04b}. The relatively large value of the sound velocity at
present produces strong oscillations and power suppression in the
matter spectrum, which render these models incompatible with current
data\footnote{A particular kind of entropy perturbations solves this
problem \cite{reis03a,reis04b} but, as discussed in section
\ref{PerfectFluid}, models with purely kinetic Lagrangians for
scalar fields are adiabatic. Even other scalar field models which
have the same background evolution as the Chaplygin gas seem to be
highly disfavored \cite{perrotta04}.} \cite{Sandvik,reis04b}. The
MBI model could avoid this problem since the sound speed is
significative only during some limited interval of densities, as
shown in the figure.

\begin{figure*}[thpb]
\centering \hspace*{0.in}
\includegraphics[height= 12 cm,width=14cm]{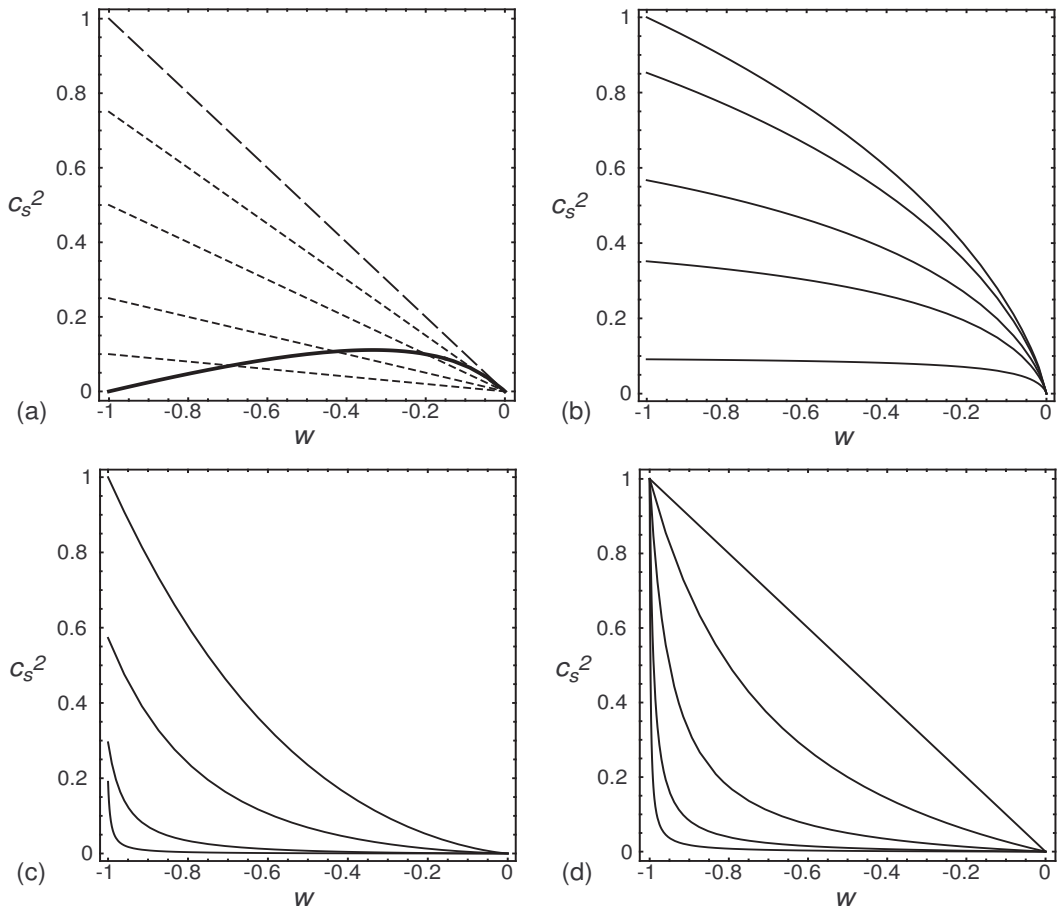}
\caption{The sound velocity $c_{s}^{2}$ as a function of the
equation of state parameter $w$ for the MBI case and other
quartessence models. (a) Results for the MBI model (thick line) and
the Generalized Chaplygin gas, with $\alpha=0.1,0.25,0.5$, and
$0.75$ (from bottom to top, short dashed curves), and $\alpha=1$
(equivalent to the Born-Infeld model, long dashed line). (b)
Exponential quartessence with $\alpha=0.1,0.5,1,2,$ and $e$ (from
bottom to top). (c) Logarithmic quartessence with $\alpha=10^{-3}
,10^{-2},10^{-1},$ and $e^{-1}$ (from bottom to top). (d) Results
for the $n^{th}$ order Dirac--Born--Infeld theory with
$n=2,5,20,100$, and $500$ (from top to bottom).} \label{cs2w}
\end{figure*}

It is useful to see how the sound velocity $c_{s}^{2}$ evolves as a
function of $w:=p/\rho,$ which is sometimes know as the ``equation
of state parameter''. Both $c_{s}^{2}$ and $w$ are the quantities
that appear in the equations for the evolution of perturbations in
the fluid representation (see e.g. ref. \cite{KodamaSasaki}). The
$c_s^2 \times w$ relation allows also to make a comparison with
other quartessence models.

In the MBI case we have $w=1-\Sigma^{2}/W$, such that
\[
c_{s}^{2}=\frac{w\left(1+w\right)  }{3w-1}\,.
\]
In the Born--Infeld model, we have $c_{s}^{2}=-w$, while for the
Generalized Chaplygin gas \cite{Kamenshchik01,Billic02,bento,GCG1}
the relation is $c_{s}^{2}=-\alpha w$. Recently, another
generalization of the Born--Infeld action was proposed, the $n^{th}$
\emph{order Dirac--Born--Infeld theory}, given by
$L\propto-(1-W)^{1/n}$ \cite{AbramoQuartessence}, in which case
$c_{s} ^{2}=-2w/(n+(n-2)(1+2w))$. Other two models of quartessence
were investigated in \cite{skewness,reis04b}, where the equations of
state are given by $p=-M^{4}\exp\left( -\alpha\rho/M^{4}\right)$ and
$p=-M^{4}/\left[\log\left(\rho/M^{4}\right)\right] ^{a}$. Here we
assume adiabatic perturbations of these models such that
$c_{s}^{2}=dp/d\rho$.

In figure \ref{cs2w} we plot $c_{s}^{2}$ as a function of $w$ for
the models referred above. In the upper left panel we show the
results for the MBI (thick line), Born--Infeld (long dashed line),
and the Generalized Chaplygin gas models (short dashed curves). In
the upper right, the results for the exponential quartessence are
displayed. In the lower left panel we plot curves for the
logarithmic quartessence model. Finally, in the lower right panel of
this figure we show curves for the $n^{th}$ \emph{order
Dirac--Born--Infeld theory}. From these figures, it is clear that
all the models start with $c_{s}^{2}=0$ at $w=0$, where quartessence
acts as CDM. Nevertheless, in the MBI case, the sound velocity
vanishes again when $w=-1$, whereas in the other models displayed in
the figure the sound velocity attains a maximum at that point. This
leads to oscillations and suppression in the power spectrum of
density perturbations.\footnote{This was shown explicitly for the
Generalized Chaplygin Gas \cite{Sandvik} and the Exponential and
Logarithmic models \cite{reis04b}. The behavior of perturbations was
not yet studied in the $n^{th}$ \emph{order Dirac--Born--Infeld
theory} \cite{AbramoQuartessence}.} Thus we expect that this issue
of the power spectrum will be alleviated in the MBI model.

\section{Dynamics of Spatially Homogeneous Solutions\label{dyna}}

Now, let us focus on the dynamics of the MBI model for spatially
homogeneous and isotropic configurations. In this case the
Raychaudhuri equation is given by\footnote{For simplicity, in the
following sections we shall consider a universe composed only of
scalaron matter (baryons will be included in sections
\ref{HubbleMBI} and \ref{SNIaMBI}). We choose units such that $8\pi
G=1$.}
\begin{equation}
\dot{\theta}+\frac{1}{3}\theta^{2}+\frac{1}{2}(\rho+3p)=0\,,
\label{4Raychaudhuri}%
\end{equation}
the energy conservation equation is
\begin{equation}
\dot{\rho}+\left(\rho+p\right)\theta=0\,, \label{econs}
\end{equation}
and the expansion parameter can be written as $\theta=3\dot{a}/a$,
where $a\left(  t\right)  $ is the scale factor.

Combining equations (\ref{4Raychaudhuri}) and (\ref{econs}) we obtain a
dynamical system:%
\begin{equation}
\frac{d\rho}{dt}=F(\rho,\theta)\,,\qquad\frac{d\theta}{dt}=G(\rho,\theta)\,,
\label{D1}%
\end{equation}
with
\begin{equation}
F=-\theta\frac{AW(2W-\Sigma^{2})\,}{\sqrt{W\left(\Sigma^{2}-W\right)}
},\;G=-\frac{1}{3}\theta^{2}-\frac{1}{2}\,\frac{AW\left(4W-3\Sigma
^{2}\right)}{\sqrt{W\left(\Sigma^{2}-W\right)}}, \label{FG}%
\end{equation}
where, from eq. (\ref{rhoMBI}), $W$ is given by
\begin{equation}
W\!=\!\frac{2^{-1/3}\!\left(  3A\Sigma^{2}{\rho}^{2}\right)^{2/3}%
\!\!{\left(1\!+\!\sqrt{1\!+\!\frac{4\,{\rho}^{2}}{27\,A^{2}\,\Sigma^{4}}%
}\right)}^{\!2/3}-\left(2/3\right)^{1/3}\!{\rho}^{2}}{\left(
9\,A^{4}\,\Sigma^{2}\,{\rho}^{2}\right)^{1/3}{\left(1\!+\!\sqrt
{1\!+\!\frac{4\,{\rho}^{2}}{27\,A^{2}\,\Sigma^{4}}}\right)}^{1/3}}.
\label{wrho}%
\end{equation}
From an inspection of (\ref{4Raychaudhuri}, \ref{econs}) it is
easy to find the four stationary points $P(\rho,\theta)$ of this
system. Two correspond to the
fundamental state $W=\Sigma^{2}/2$,%
\begin{equation}
M=(\rho_{\mathrm{v}},\sqrt{3\rho_{\mathrm{v}}})\,,\qquad
N=(\rho_{\mathrm{v}},-\sqrt{3\rho_{\mathrm{v}}})\,, \label{D3}
\end{equation}
and two are on the line $\theta=0$,
\begin{equation}
O=(0,0)\,,\qquad Q=(\rho_{c},0)\,,\label{D2}%
\end{equation}
where $\rho_{\mathrm{v}}$ is given by (\ref{rhov}) and $\rho_{c}=
(3\sqrt{3}/2)\rho_{\mathrm{v}}$. Consideration of the behavior of
the system in the neighborhood of these points shows that $M$ and
$N$ are two-tangent nodes (one stable and the other unstable) while
those on the $\theta=0$ line are saddle points.

\begin{figure}[ptb]
\centering \hspace*{0.in}
\includegraphics[height= 9.4 cm,width=8.7 cm]{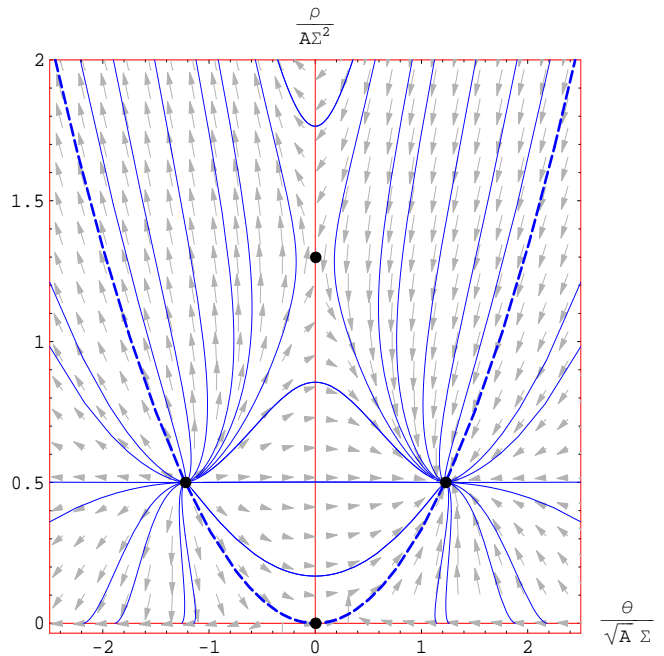}
\caption{Vector field and trajectories of the MBI model in the
$(\rho,\theta)$ plane. The dashed curve is the trajectory for a
flat universe. The dots correspond to the four stationary points
$M,N,Q$, and $O$ discussed in the
text.}%
\label{trajMBI}%
\end{figure}

In figure \ref{trajMBI} we display the phase diagram in the
$\left(\rho,\theta\right)$ plane for equations
(\ref{D1}-\ref{wrho}), as well as some trajectories. From expression
(\ref{rhoMBI}) we see that only positive energy-density solutions
are allowed. From the figure, it is clear that any trajectory that
starts with a density lower than the fundamental state will remain
with $\rho\leq \rho_{\mathrm{v}}$. Conversely, trajectories starting
at $\rho>\rho_{\mathrm{v}}$ are bounded to have
$\rho\geq\rho_{\mathrm{v}}$. It is easy to see from the Friedmann
equation that the solutions for the flat case $(k=0)$ give
$\rho=\theta^{2}/3$, which is the dashed trajectory plotted in this
figure. Notice the existence of a trajectory with vanishing density
and nonzero expansion (the bottom horizontal line). This is the well
known Milne solution, and is allowed only if we have $e=0$ in
Lagrangian (\ref{EBILag}), leading to our nomenclature for this
model.

From now on, we shall be concerned with the right hand side of
figure \ref{trajMBI}, i.e., with expanding solutions. Some of
these may correspond to the background (unperturbed) cosmological
solution, in particular those starting from high values of $\rho$.
As a nearly flat spatial geometry is favored by current
astrophysical observations, one should only consider trajectories
near the $k=0$ one. From figure \ref{trajMBI} it is clear that
these trajectories end up in the fundamental state $M$. In other
words, \emph{the de Sitter solution is an attractor for the
trajectories of cosmological interest}.

It is important to notice that, even in the inhomogeneous case,
there is a ``vacuum barrier'' that cannot be transposed in any
expanding region. This can be seen from the energy conservation
equation (\ref{econs}), which is valid for any $\theta$. As the
vacuum state $p=-\rho$ is approached, the density freezes out, and
this is a stable point (the contrary occurs in a contracting
region). In the MBI model, only states that initially violate the
weak energy condition are able to approach the point $O$, which
represents a ``static vacuum'' ($\varphi=const.$). Therefore, as
pointed out before, only the domain between the dust regime
$W\rightarrow\Sigma^{2}$ and the fundamental state
$W_{0}=\Sigma^{2}/2$ --- which represents a ``dynamical vacuum''
--- is relevant for the cosmological model. We shall see below
that, within this interval, the universe has an accelerated
expansion phase.

\section{The Acceleration of the Universe\label{MBIaccel}}

The acceleration of the universe is controlled not only by the energy density
but also by the pressure, according to the Raychaudhuri equation
(\ref{4Raychaudhuri}). Writing this equation in terms of the scale factor $a$
and using equations (\ref{rhoMBI}) and (\ref{pMBI}) yields:
\begin{equation}
\frac{\ddot{a}}{a}=\frac{AW\left(  4W-3\Sigma^{2}\right)  }{\sqrt{W\left(
\Sigma^{2}-W\right)  }} \,. \label{A1}%
\end{equation}
Starting with $W$ close to $\Sigma^{2}$, that is, a matter dominated
universe,
it is clear that the acceleration changes sign at%
\begin{equation}
W_{a}=\frac{3}{4}\,\Sigma^{2}. \label{A3}%
\end{equation}
If the scalaron is the dominant component of the cosmological
energy density, the universe will begin to accelerate when $W$ is
close to $W_{a}$. Therefore, along the transition from a dust
universe to a de Sitter one, the expansion starts accelerating, as
expected. Close to this transition point the sound velocity
reaches its maximum value $c_{s}^{2}=1/9$, as discussed in section
\ref{stability}.

\section{The Hubble parameter\label{HubbleMBI}}

As discussed in the preceding sections, the MBI Lagrangian appears
as a interesting model for the matter content of the universe,
providing an explanation for the accelerated expansion as well as
for structure formation. Clearly, in the cosmological setting, among
all solutions displayed in the phase plane of figure \ref{trajMBI}
we must concentrate in the domain characterized by the inequality
$\Sigma^{2}/2<W<\Sigma^{2}$.

To compare the predictions of the model to observational data on
the background dynamics, such as the redshift-distance relation,
we need to know the evolution of the Hubble parameter. For this
sake, the energy density has to be expressed in terms of the scale
factor. This is accomplished by solving the energy conservation
equation (\ref{econs}), which is conveniently expressed in terms
of $W$. Using (\ref{rhoMBI}) and after a few manipulations,
we get%
\[
-3\frac{\dot{a}}{a}=\dot{W}\left(  \frac{1}{2}\frac{1}{\Sigma^{2}-W}-\frac
{2}{\Sigma^{2}-2W}\right)  .
\]
Integrating the above equation, in the desired range of $W$, we
obtain
\begin{equation}
\Sigma\sqrt{8C}a^{-3}=\frac{2W-\Sigma^{2}}{\sqrt{\Sigma^{2}-W}}\,,
\label{solEconsMBI}%
\end{equation}
where $C$ is a (conveniently defined) integration constant. This
relation is
easily inverted to get%
\begin{equation}
W=\left[  \frac{1}{2}+Ca^{-6}\left(  \sqrt{1+C^{-1}a^{6}}-1\right)  \right]
\Sigma^{2}, \label{WaMBI}%
\end{equation}
were the sign of the root was chosen such that
$W\!\rightarrow\!\Sigma^{2}$ when $a\!\rightarrow\!0$, to have a
dust behavior in the young universe. Inserting (\ref{WaMBI}) in
(\ref{rhoMBI}) and using (\ref{solEconsMBI}), we finally get%
\begin{equation}
\rho=A\Sigma^{2}\sqrt{\frac{2}{C}}\frac{\left[  \frac{1}{2}+Ca^{-6}\left(
\sqrt{1+C^{-1}a^{6}}-1\right)  \right]  ^{3/2}}{\sqrt{1+C^{-1}a^{6}}-1}%
\,a^{3}. \label{rhoaMBI}%
\end{equation}
It is easy to see that, in the limit $a\!\rightarrow\!0$,
$\rho\!\rightarrow \!A\Sigma^{2}\sqrt{8C}a^{-3}$, and the scalaron
density scales as CDM. On the other hand, in the limit
$a\!\rightarrow\!\infty$, the fundamental state (\ref{rhov}) is
reached. Thus, as expected, this model behaves as CDM at early times
and goes to a cosmological constant-like state in the future, being
a candidate for quartessence. In fact, this asymptotic behavior
occurs in any equation of state that intersects the $p=-\rho$ line
\cite{tese}.

In a realistic cosmological situation one has to consider also the baryons and
the radiation components. Defining $\Omega_{i}=\rho_{i}/\rho_{crit}=8\pi
G\rho_{i}/\left(  3H^{2}\right)  $, with $H=\dot{a}/a$, the Friedmann equation
is given by%
\[
H^{2}=\frac{8\pi
G}{3}\sum_{i}\rho_{i}-\frac{k}{a^{2}}=H_{0}^{2}\left(
\sum\Omega_{i}^{0}\frac{\rho_{i}}{\rho_{i}^{0}}+\Omega_{k}^{0}a^{-2}\right)
\,,
\]
where $\Omega_{k}^{0}=-k/H_{0}^{2}$, the subscript $i$ denotes each
component, and the knot means that the quantity is computed at the
present value of the scale factor $a=a_{0}:=1$. For baryons
$\rho_{b}/\rho_{b}^{0}=a^{-3}$, while for radiation
$\rho_{r}/\rho_{r}^{0}=a^{-4}$. The energy density in the scalaron
field (\ref{rhoaMBI}) shall be denoted as $\rho_{s}$. From
expression (\ref{rhoaMBI}) it is clear that $\rho_{s}/\rho_{s}^{0}$
is a function only of $C$ and $a$. Therefore, as in the BI case,
although the Lagrangian (\ref{MBILag}) has two free parameters,
\emph{the observables depend upon a single free parameter}. As the
parameter $C$ lacks a direct physical interpretation, it is
convenient to define the quantity
\begin{equation}
R:=\frac{\rho_{\mathrm{v}}}{\rho_{s}^{0}}=\frac{A\Sigma^{2}}{2\rho_{s}^{0}}\,,
\label{RMBI}%
\end{equation}
which, as can be seen from (\ref{rhoaMBI}), depends only on $C$. The
parameter $R$ is a measure of how close the present energy density
in the scalaron field is to its final ``vacuum'' value. In other
words it is a measure of how close our universe is to its final de
Sitter state. From the definition above, it is clear that $0\leq
R\leq1$. The explicit form of $C\left(  R\right)  $ is%
\begin{equation}
C=\frac{1}{8}\frac{\left(  2f-1\right)  ^{2}}{1-f}\,, \label{CR}%
\end{equation}
with%
\begin{equation}
f=\frac{\left(  3\sqrt{3}R+\sqrt{1+27R^{2}}\right)  ^{2/3}-1}{R\left(
3\sqrt{3}R+\sqrt{1+27R^{2}}\right)  ^{1/3}}\frac{\sqrt{3}}{6}\,. \label{fR}%
\end{equation}

Defining $\rho_{s}/\rho_{s}^{0}:=g\left(a,R\right)$, were $g$ is
obtained from (\ref{rhoaMBI}) together with (\ref{CR}) and
(\ref{fR}), the Hubble function will be
given by%
\begin{equation}
\frac{H^{2}}{H_{0}^{2}}=\Omega_{s}^{0}g\left(  a,R\right)  +\Omega_{b}%
^{0}a^{-3}+\Omega_{r}^{0}a^{-4}+\Omega_{k}^{0}a^{-2}\,, \label{H2}%
\end{equation}
where $\Omega_{k}^{0}=1-\left(  \Omega_{s}^{0}+\Omega_{b}^{0}+\Omega_{r}%
^{0}\right)$ and $H_{0}$ is the Hubble constant. In the next section
we shall use recent data on the distance-redshift relation
determined from type Ia supernova observations to set constraints on
$R$.

\section{Constraints from Type Ia Supernovae\label{SNIaMBI}}

Type Ia supernovae (SNIa) observations have now become a standard
tool for cosmology. Empirical studies show that these supernovae
can be good standard candles after light curve calibration (see
e.g. refs \cite{SNreview,recentSNIa}). The first compelling
evidence from a single experiment for the accelerated expansion of
the universe was indeed based on SNIa data \cite{discoverySNIa}.
Larger data sets including higher redshift supernova are currently
available \cite{tonry03,Barris04,HSTSNIa}. These new data makes a
stronger case for the current accelerated expansion, since, as
expected, a deceleration is detected for higher redshift
supernovae, weakening the concerns that the reported acceleration
could be due to systematic effects.

Here we shall use some of these recent data to check whether the
MBI model can provide a good fit to SNIa observations and to set
constraints on the parameter $R$ (eq. \ref{RMBI}). For this sake,
a sample of 253 SNIa was compiled from references \cite{Barris04}
and \cite{tonry03} which provide tables with the redshift $z$,
$\log D_{L}$, and its variance $\sigma_{\log D_{L}}^{2}$ for each
supernovae. Here $D_{L}$ is the luminosity distance times the
Hubble constant, which, in the flat case, is given by
\begin{equation}
D_{L}=d_{L}H_{0}=(1+z)\,H_{0}\int_{0}^{z}\frac{dz^{\prime}}{H(z^{\prime})}\,.
\label{DL}%
\end{equation}
The Hubble parameter $H\left(z\right)$ is obtained from equation
(eq. \ref{H2}) with $z=a^{-1}-1$.

Since observations of anisotropies in the CMBR indicate that the
Universe is nearly flat ($\Omega_{k}\ll1$), the analysis that
follows will be restricted to the zero curvature case. Also, at the
redshifts probed by SNIa data, the radiation contribution to the
Hubble parameter (eq. \ref{H2}) is negligible. Therefore, we set
$\Omega_{s}^{0}+\Omega_{b}^{0}=1$. We also fix the baryon density
parameter at $\Omega_{b}=0.04$, in agreement with the observed
abundances of light elements \cite{BBN,DH} together the Hubble Space
Telescope key project result \cite{freedman}. After these choices,
$D_{L}$ becomes solely a function of the MBI parameter $R$.

Following \cite{Barris04} and \cite{tonry03} we discard local
supernovae with $z<0.01$, because the peculiar motion contribution
to $z$ is too high, and those with high host extinction,
$A_{V}>0.5$, which could cause a strong bias in the determination of
$D_{L}$. After these cuts, we end up with a sample of $194$ SNIa
extending up to $z=1.75$. Despite eliminating low redshift SNIa,
peculiar motions affect the measurement of $z$ at all redshifts,
causing a scatter around the value given by the bulk cosmic
expansion. This is taken into account by including an uncertainty in
$z$ which is propagated into the luminosity distance and is summed
in quadrature with the observational uncertainty in $D_{L}$.
Therefore, the Chi-squared is defined by:
\[
\chi^{2}=\sum_{i=1}^{194}\frac{\left[  \log\left(  D_{L}^{Obs}\left(
z_{i}\right)\right)-\log\left(D_{L}^{Th}\left(z_{i}\right)\right)
\right]^{2}}{\sigma_{\log\left(  D_{L}\left(z_{i}\right)\right)}
^{2}+\left(\left.\frac{\partial\log D_{L}^{Th}}{\partial
z}\right\vert _{z_{i}}\sigma_{z}\right)^{2}}\,,
\]
where the theoretical prediction is computed from (\ref{DL})
together with (\ref{H2}) and the observational values are given in
the tables of \cite{Barris04} and \cite{tonry03}. Here we assume a
velocity dispersion of $\sigma_{v}=500Km/s$ (such that
$\sigma_{z}\simeq\sigma_{v}/c\simeq 1.7\times 10^{-5}$).

With the above described assumptions and data set, the best fit
value is $R=0.713$ which corresponds to $\chi^{2}=198.0$.
Therefore the Chi-squared per effective degrees of freedom ($193$
in this case) is given by $\chi _{dof}^{2}=1.026$. The MBI model
provides a fit as good as for both the $\Lambda$CDM
($\chi_{dof}^{2}=1.029$) and Born--Infeld ($\chi_{dof}^{2}=1.020$)
models, which also have a single free parameter. The reduced
Chi-squared is also very close to other models of quartessence
with two free parameters, such as the Generalized Chaplygin Gas
($\chi_{dof}^{2}=1.025$), Exponential Quartessence
($\chi_{dof}^{2}=1.021$), and Logarithmic Quartessence ($\chi_{dof}^{2}%
=1.023$) \cite{CMBRpheno}. However, as pointed out before, these models are
incompatible with large-scale structure data for adiabatic perturbations
\cite{Sandvik,reis04b}.

Defining the likelihood by $\mathcal{L}(R)\propto
e^{-\chi^{2}/2}$, we have checked that the probability
distribution of $R$ is very close to a Gaussian, with less than
$1\%$ deviation. At $1\sigma$ ($68.3\%$) confidence level we have
$R=0.713\pm 0.029$. Thus, the current SNIa data is able to limit
the MBI free parameter to a rather narrow interval. This makes the
model readily testable by using different observables. If the
values of $R$ are incompatible to the ones obtained from SNIa, the
model must be discarded.

\section{Concluding Remarks\label{conclusion}}

Inspired in the Born--Infeld theory for a scalar field, we have
introduced a new model for dark-matter and dark-energy
unification, the EBI Lagrangian (\ref{EBILag}). We have developed
a systematic study of a subclass of this Lagrangian, namely the
MBI model, for which the trajectories in the phase space of
homogeneous cosmologies were computed. It was shown that the
trajectories of cosmological relevance end up asymptotically in a
de Sitter state.

A new feature of this model is the fact that the equation of state
changes concavity. Therefore, the sound velocity is maximal at an
earlier epoch, contrary to the previously considered convex
quartessence models, where $c_s$ is greater at present
\cite{reis04b}. This could solve the known problems associated
with the power spectrum of adiabatic quartessence \cite{Sandvik}.
Therefore, this model has advantages over other generalizations of
the Born--Infeld Lagrangian \cite{bento,AbramoQuartessence}.

We have obtained an explicit form for the Hubble function, which
depends on a single free parameter (once the $\Omega_i^0$ are
fixed), expressed in terms of the ratio of the fundamental state
and current scalaron densities
($R=\rho_{\mathrm{v}}/\rho_{s}^{0}$). We applied this result to
obtain the luminosity distance-redshift relation and used recent
type Ia supernovae data to place constraints on $R$ assuming a
flat universe. The model provides a good fit to the data, giving a
similar $\chi^2_{dof}$ as the $\Lambda$CDM model --- with the same
number of degrees of freedom
--- and other models of quartessence.

We find that $R=0.713\pm 0.029$, at $68.3\%$ confidence.
Therefore, contrary to other quartessence models, whose allowed
parameter values are highly degenerate
\cite{GCG1,GCG2,GCG4ess,CMBRpheno}, this model is tightly
constrained by current data. Once $R$ is fixed, the model provides
clear predictions for other cosmological observables, such as the
matter and CMBR power spectrum.

The results reported in this paper motivate further studies of the
EBI model, both to perform a detailed study of its dynamics in the
general case, as well as to obtain observational constraints on the
model. In particular, in what regards the MBI case, a more complete
observational analysis should be performed, both to check if the
model withstands to other observables and to improve the statistical
analysis. For example, we could have marginalized over the Hubble
constant and baryon fraction, instead of fixing these parameters to
their best-fitting values. We could also study the effects of
spacial curvature on the determination of the parameters. It is
fundamental to investigate the formation of cosmic structures in
this scenario, both in the linear and nonlinear regimes.
Nonetheless, the results of this work indicate that the model here
introduced may be a viable candidate for quartessence.

As a final remark, let us recall that in the quartessence model
there is an effective transformation of dark-matter into dark-energy
and that, in the homogeneous case, around $96\%$ of the cosmic
matter ends up in the de Sitter fundamental state. This brings us to
a comment by S. Carrol, stating that ``[the de Sitter solution]
represents the only natural stable solution for cosmology, and one
of the outstanding problems is why we don't find ourselves living
there'' \cite{Carrolllectures}. In the MBI model, and in the
quartessence scenario in general, the answer to this question is: we
are indeed heading there!

\begin{acknowledgments}

We thank the participants of the ``Pequeno Semin\'{a}rio'' for
useful comments and suggestions. MM acknowledges Raul Abramo,
Patrick Greene, Francesca Perrotta, and Ja\'{\i}lson Alcaniz for
useful comments and suggestions. MM is grateful to the organizers of
the COSMO'04 meeting and wishes to thank the hospitality of the
Theoretical Astrophysics Group at Fermilab. MN is partially
supported by the Brazilian research agencies CNPq and FAPERJ. MM
acknowledges financial support from FAPERJ and CNPq. LSW is
partially supported by FUJB. CAR is partially supported by CNPq.
\end{acknowledgments}

\appendix*

\section{Speculations on the Possible Role of Scalarons in the Early
Universe\label{ApTachyon}}

In this section, we conjecture about the possible role of
scalarons in solving two fundamental problems of cosmology: the
homogenization of the universe and the issue of initial
conditions. First, let us point out that scalarons propagate as
tachyons. Indeed, from section \ref{EffGeo} the propagation of the
scalarons is provided by $k_{\mu}\,k_{\nu}\,g_{eff}^{\mu\nu}=0$,
where the effective metric is given by equation (\ref{C4}). Thus,
in the domain $\Sigma^{2}>W>\Sigma^{2}/2$ we find, from
(\ref{kk}), that the scalarons propagate outside the local light
cone
\begin{equation}
k_{\mu}k_{\nu}g^{\mu\nu}=\frac{\Sigma^{4}}{\Sigma^{2}-2W}\,\frac{(k_{\mu
}\nabla^{\mu}\varphi)^{2}}{L^{2}}<0\,. \label{U2}%
\end{equation}
For the present energies in the cosmos, the interaction of the
scalaron with normal matter must be very weak, otherwise it would
not be dark or it would have already been detected in the
laboratory. However, the coupling of the scalaron to ordinary
matter could be important in the early universe. One may wonder if
the superluminal propagation of scalarons could have an important
role in the homogenization of the universe.

Another interesting point relates to the initial conditions. One
of the possible natural initial configurations is that the scalar
field be produced in the state of maximal ``kinetic term'', i.e.,
with maximal space-time fluctuations. If on one hand it may be
hard to understand how could the universe be ``prepared'' in a
state of infinite density, on the other hand it is perhaps natural
to accept that the scalaron was ``prepared'' in such a state of
maximal kinetic term
$\nabla_{\mu}\varphi\,\nabla^{\mu}\varphi=\Sigma^{2}$, which in
turn leads to a divergent $\rho$. If radiation is coupled to the
scalaron in the early universe, the temperature could be very high
due to the thermal equilibrium with the scalaron. A coupling to
the electromagnetic field could be introduced in a simple way
considering a gauge interaction with a complex scalar field.
Notice that the results of the previous sections do not change if
we consider a complex scalar field instead of a real one. For a
discussion of couplings to other fields see e.g. refs.
\cite{GhostCond,Fierz}.

Therefore, the MBI model might provide some hints about the early
universe, being responsible for the extremely high densities as well
as for the cosmic homogenization. These speculations will be subject
of further study.

\end{document}